\documentclass[a4paper,fleqn]{cas-dc}
\usepackage[natbibapa]{apacite}
\AtBeginDocument{%
  \renewcommand{\APACrefDOI}[2]{%
    \ifx\@empty#2\@empty
      {}%
    \else
      \href{https://doi.org/#2}{#2}%
    \fi
  }%
}

\usepackage[acronym,toc]{glossaries}
\newacronym{stem}{STEM}{Science, Technology, Engineering, and Mathematics}
\newacronym{apa}{APA}{American Psychological Association}
\newacronym{lms}{LMS}{Learning Management System}
\newacronym{ioc}{IoC}{Internationalisation of the Curriculum}
\newacronym{unesco}{UNESCO}{United Nations Educational, Scientific and Cultural Organisation}
\newacronym{pal}{PAL}{Personalised and Adaptive Learning}
\newacronym{its}{ITS}{Intelligent Tutoring Systems}
\newacronym{ai}{AI}{Artificial Intelligence}
\newacronym{ml}{ML}{Machine Learning}
\newacronym{nlp}{NLP}{Natural Language Processing}
\newacronym{genai}{GenAI}{Generative AI}
\newacronym{llm}{LLMs}{Large Language Models}
\newacronym{aied}{AIEd}{AI in Education}
\newacronym{als}{ALS}{Adaptive Learning Systems}
\newacronym{nuc}{NUC}{National Universities Commission}
\newacronym{noun}{NOUN}{National Open University of Nigeria}
\newacronym{bleu}{BLEU}{Bilingual Evaluation Understudy}
\newacronym{rag}{RAG}{Retrieval-Augmented Generation}

\usepackage[english]{babel}
\usepackage[nottoc]{tocbibind}
\usepackage{subcaption,graphicx}
\usepackage{lipsum}
\usepackage{algorithm}
\usepackage{algpseudocode}
\usepackage{orcidlink}
\usepackage{booktabs}
\usepackage{multirow}
\usepackage{siunitx}
\usepackage{geometry}
\usepackage{pdflscape}
\usepackage{array}
\usepackage{appendix}
\usepackage{afterpage}
\usepackage{graphicx}
\usepackage{subcaption}
\usepackage{float} 
\usepackage{caption} 
\usepackage{tabularx}
\usepackage{tikz}
\usepackage{pgfplots}
\usepackage{listings}
\usepackage{xcolor}

\begin{document}
\let\WriteBookmarks\relax
\def\floatpagepagefraction{1}
\def\textpagefraction{.001}

\shorttitle{An AI-Based Adaptive Learning Platform for Multilingual and Low-Resource Educational Contexts: A Case Study on Nigeria}

\shortauthors{Everistus et~al.}

\title [mode = title]{An AI-Based Adaptive Learning Platform for Multilingual and Low-Resource Educational Contexts: A Case Study on Nigeria}                      

\tnotetext[1]{This research was supported by the British Council Funded Project - STEM 4.0: Advancing Technology Education Through AI-Driven And Adaptive Learning (Project Number: TNE2024-057) and the
Department of Computer Science, Nottingham Trent University.}

%

\author[1]{Everistus Ugochukwu Nwogo\orcidlink{0009-0008-1318-7747}}
\cormark[1]
\fnmark[]
\ead{everistus.nwogo2024@my.ntu.ac.uk}

\author[1]{Isibor Kennedy Ihianle\orcidlink{0000-0001-7445-8573}}
\cormark[1]
\fnmark[]
\ead{isibor.ihianle@ntu.ac.uk}


\affiliation[1]{organization={Nottingham Trent University},
    city={Nottingham},
    postcode={NG11 8NS}, 
    country={United Kingdom}}

\author[1]{Pedro Machado\orcidlink{0000-0003-1760-3871}}
\fnmark[]
\ead{pedro.machado@ntu.ac.uk}

\author[1]{Jordan J. Bird\orcidlink{0000-0002-9858-1231}}
\fnmark[]
\ead{jordan.bird@ntu.ac.uk}

\author[1]{Ahmad Lotfi\orcidlink{0000-0002-5139-6565}}
\fnmark[]
\ead{ahmad.lotfi.ac.uk}

\author[2]{Ahmad Abdulnasir Shuaib\orcidlink{0009-0006-3791-3569}}
\fnmark[]
\ead{ahmadabdulnasir9@gmail.com}

\author[2]{Isaac Ibukun Akinwumi\orcidlink{0000-0003-3053-8189}}
\fnmark[]
\ead{isaac.akinwumi@covenantuniversity.edu.ng}

\author[2]{Jonathan Oluranti\orcidlink{0000-0002-3279-9895}}
\fnmark[]
\ead{jonathan.oluranti@covenantuniversity.edu.ng}

\affiliation[2]{organization={Covenant University},
    addressline={Km. 10 Idiroko Road,  Canaan Land,}, 
    city={Ota},
    state={Ogun State},
    country={Nigeria}}

\cortext[cor1]{Corresponding author}

\begin{abstract}
Educational platforms in under-resourced and multilingual contexts, such as Nigeria, often struggle with limited personalisation, inadequate language support, and weak curriculum internationalisation, leading to reduced learner engagement and inclusivity. This paper presents an AI-based adaptive learning platform designed for multilingual and low-resource educational contexts, with a case study on Nigerian Pidgin English. The system integrates fine-tuned large language models (LLMs) within a personalised and adaptive learning (PAL) framework, addressing linguistic inclusivity and computational constraints in resource-limited environments. To enhance linguistic alignment, a curated Nigerian Pidgin corpus was developed and used to fine-tune an instruction-tuned LLM. The study further investigates model optimisation through multi-level quantisation (4-bit, 5-bit, and 8-bit), enabling systematic analysis of trade-offs between semantic fidelity and computational efficiency. Experimental evaluation combines automatic semantic metrics (BLEU, ROUGE-L, BERTScore, perplexity, lexical diversity) with human-centred cultural assessment conducted by native speakers. Results demonstrate that higher-bit quantisation improves semantic preservation and structural coherence, while lower-bit models offer reduced inference latency with minimal degradation in instructional quality. The findings establish a deployable, resource-aware intelligent learning system that balances semantic robustness, cultural relevance, and computational efficiency. This work contributes an experimentally validated framework for adapting large language models to low-resource languages while maintaining practical feasibility for scalable educational deployment.
\end{abstract}


\maketitle

\section{Introduction}\label{sec1}
Rapid technological advancements have significantly reshaped global education, particularly in the way information is delivered, stored, and accessed \citep{Essa}. Although digital innovations such as e-learning platforms promise improved accessibility, flexibility, and scalability, their benefits remain unevenly distributed, especially in multilingual and resource-constrained educational contexts. In such settings, structural inequalities, linguistic barriers, and limited digital infrastructure continue to restrict meaningful access to quality education. According to \gls{unesco}, globally, approximately 91\% of boys and 89\% of girls of primary school age are enrolled annually, yet only 88\% progress to junior secondary education, and approximately 78\% reach senior secondary levels ({\href{https://ourworldindata.org/data-insights/a-century-of-progress-in-access-to-primary-education}{Our World in Data}; \href{https://www.unesco.org/gem-report/en/view/completion}{UNESCO}}).

At the tertiary level, enrollment reached 264 million in 2024, but only about 43\% of students completed their programs on time (\href{https://www.unesco.org/en/higher-education}{UNESCO}; \href{https://www.oecd.org/en/publications/education-at-a-glance-2025_1c0d9c79-en.htmlk}{OECD}). Globally, \gls{stem} graduates constitute less than 25\% of university graduates, with lower representation in low- and middle-income countries (\href{https://unesdoc.unesco.org/ark:/48223/pf0000377250}{UNESCO}). These figures highlight persistent challenges in ensuring equitable educational opportunities, particularly in fields critical to technological and economic development.

E-learning supported by \gls{lms} platforms has emerged as a key innovation to address persistent educational gaps. These platforms integrate content delivery, assessment, and learner support within a unified digital environment \citep{Simanullang, Bradley}. A recent systematic review of generative AI in education further emphasises the lack of standardised technological evaluation frameworks and raises concerns about trustworthiness and contextual reliability \citep{AI_Future_of_Education_Review}. However, limited work addresses deployment under multilingual and infrastructural constraints, particularly in low-resource settings. Traditional classroom models typically rely on uniform instructional approaches that limit personalisation and often do not align with individual learning styles, leading to reduced engagement and suboptimal learning outcomes \citep{Sharma, Shirazi, Magfirah}. More recently, AI-driven educational technologies have been proposed as a means to overcome the limitations of traditional LMS platforms. However, many AI-based learning systems are developed and evaluated in high computational-resource, monolingual contexts, implicitly assuming reliable connectivity, computational capacity, and linguistic homogeneity. As a result, these systems often struggle to generalise to multilingual classrooms and low-resource environments, where language diversity, infrastructural instability, and cultural context play a critical role in learning effectiveness. Similarly, conventional summative assessment practices focus primarily on the outcomes rather than the learning process itself, providing limited insight on learner progress or lack thereof \citep{Double, Meylani}. As education increasingly shifts towards personalised and learner-centred models, formative assessment and continuous feedback have become essential for identifying individual learning gaps and supporting adaptive learning pathways \citep{Yang, Elbanna}.

Linguistic and cultural barriers further constrain learning, particularly in multilingual contexts such as Nigeria. Although English is widely used as an instruction medium, it is neither the first language nor the lingua franca of many learners \citep{Akujobi}. Although \gls{unesco} advocates mother-tongue instruction to improve comprehension and learning outcomes, the dominance of English-only instruction in Nigeria often marginalises students and exacerbates educational inequities \citep{Onuoha}. Within this multilingual landscape, Nigerian Pidgin English plays a unique role as an informal lingua franca across ethnic, regional, and educational boundaries, particularly in urban and semi-urban contexts. Unlike many indigenous languages that are regionally bounded, Nigerian Pidgin is widely used in everyday communication, popular media, and peer-to-peer learning interactions. Its broad social reach and functional flexibility make it a pragmatic and inclusive target for educational AI systems aimed at reducing linguistic barriers and improving learner engagement in Nigeria.

Beyond linguistic challenges, Nigeria’s education system is characterised by significant structural and resource constraints. The country has more than 520 indigenous languages and over 20 million out-of-school children, the highest number globally, highlighting persistent limitations in both access and educational quality. Rural and underserved communities face acute shortages of qualified teachers, instructional materials, and digital infrastructure. In addition, outdated curricula often fail to incorporate global competencies and the principles of \gls{ioc}, further constraining learners’ preparedness for an increasingly interconnected and technology-driven world \citep{Bello, Clifford, akinsola}.

Given these barriers, \gls{pal} technologies have gained global attention for their ability to tailor instruction to learners’ abilities, behaviours, and progress, providing support comparable to one-to-one tutoring \citep{Peng, Essa, Shemshack, Muangprathub}. Advances in \gls{ai}, including \gls{ml}, \gls{nlp}, and \gls{genai}, further strengthen \gls{pal}, enabling adaptive feedback, analytics, and personalised content delivery \citep{Zawacki, Rane}. These innovations help address teacher shortages and curriculum inconsistencies in regions such as sub-Saharan Africa \citep{Ewing, Sajja}. However, implementation of \gls{pal} in Nigeria remains limited due to lack of infrastructure, inconsistent curriculum delivery, language diversity, and socio-economic inequalities \citep{Eli, Bubou}. Many systems are designed for well-resourced contexts and do not align with Nigeria’s realities, including unreliable electricity, weak internet connectivity, and multilingual classrooms. A lack of localised research further limits the development of culturally and linguistically appropriate solutions \citep{Ayeni}. These observations reinforce the need for empirically validated and deployable intelligent learning architectures.

These challenges highlight the urgent need to contextualise and personalise AI-driven educational technologies for Nigeria’s diverse population. Without such adaptation, the transformative potential of \gls{pal} will remain inaccessible to learners who could benefit the most. In response to these open issues, the novel scientific contributions of this study are as follows:
\begin{itemize}
\item The design and implementation of an LLM-based personalised and adaptive learning architecture tailored to multilingual and low-resource educational environments, explicitly accounting for linguistic diversity, infrastructural limitations, and curriculum localisation within the Nigerian education context.
\item Fine-tuning of large language models on a curated Nigerian Pidgin English corpus and integration of Retrieval-Augmented Generation to enhance linguistic alignment, contextual grounding, and assessment reliability.
\item A systematic empirical evaluation of multi-level model quantisation (4-bit, 5-bit, and 8-bit), analysing trade-offs between semantic fidelity, fluency, and computational efficiency to support deployable intelligent system design.
\item Integration of cognitive and selected non-cognitive learner attributes within the adaptive framework to enable personalised feedback and dynamic assessment, contributing to the scalable design of inclusive AI-driven educational systems.
\end{itemize}

The structure of this paper is as follows: Section \ref{sec2} presents and analyses the related work. An overview of the field and the research questions that arise are described in Section \ref{sec3}, followed by the proposed methodology in Section \ref{sec4}. Section \ref{sec5} describes the system design and technical architecture, while Section \ref{sec6} details the experimental methodology. The results and discussion are presented in Section \ref{sec7}, and finally, conclusions and directions for future work are provided in Section \ref{sec8}.

\section{Related Work}\label{sec2}
\subsection{The Evolution of AI in Education}

The development of \gls{aied} has progressed from early rule-based systems to today’s data-driven and generative models. \cite{Guan} traced the foundations of \gls{aied} to the 1970s, building on pioneering systems such as ELIZA (1966), SCHOLAR (1970), and the expert system MYCIN (mid-1970s), which significantly influenced the design of early \gls{its}.

According to \cite{Williamson}, the broader concept of \gls{ai} emerged from the Dartmouth workshop in 1956, marking the formal beginning of the field. \gls{aied} gained institutional structure with the launch of the \textit{``International Journal of Artificial Intelligence in Education''} in 1989 and the establishment of the \textit{``International AIEd Society''} in 1993. Since then, research has increasingly focused on personalisation, learner modelling, and intelligent instructional support through \gls{its}. As noted by \cite{Kelkar}, \gls{its} evolved from the developments of Expert Systems in the 1960s and 1970s. Following the AI winter of the 1980s, the field shifted toward integrating AI with learning sciences, culminating in the Cognitive Tutor developed in the 1980s–90s, which later gained widespread adoption in US schools.

Advances in \gls{nlp} have further shaped the trajectory of \gls{aied}. \cite{Khensous} details key milestones that range from early machine translation efforts in 1954 and Chomsky’s transformational grammar (1957) to influential systems such as ELIZA (1966), SHRDLU (1970), and LUNAR (1978). Modern breakthroughs accelerated with voice assistants like Apple’s Siri (2011) and recent deep learning models capable of handling linguistic complexity and ambiguity. \cite{Chiu} highlights AI’s growing impact across learning, teaching, assessment, and administration, noting the role of tools such as ChatGPT in supporting personalised instruction, automated assessment, predictive analytics, and enhanced educational management.

In general, \gls{aied} has evolved from early dialogue-based experiments to more sophisticated adaptive systems powered by \gls{ml} and \gls{nlp}. These advances have enabled increasingly personalised learning experiences while transforming instructional practices, assessment processes, and educational administration.

\subsection{Adaptive Learning: Approaches and Outcomes}
Adaptive learning leverages data-driven strategies and AI technologies to tailor educational content and instruction to individual learners' needs, preferences, and performance. \cite{Martin} reviewed over 60 studies (2009–2018), focusing on learner, instructional, and content models. Most studies focused on cognitive outcomes (67.2\%), with some also addressing affective (37.7\%) and behavioural (41\%) aspects. Experimental methods, surveys, and online data were common. The authors proposed a framework with four key components: the learner, content, and tutoring models, and an adaptive engine, but noted its lack of clear definitions for adaptive targets and outcomes, limiting practical use.

\cite{Kabudi} analysed 147 studies on AI-based learning systems, finding \gls{als} and \gls{its} as dominant technologies. Their clustering revealed five themes: System, Literature, Algorithms, Evaluation, and Framework, but they noted that many systems remain experimental.

\cite{Kolekar} developed a rule-based system for an introductory Android course with multi-format content, quizzes, and immediate feedback. User evaluations reported issues with readability and customisation, but demonstrated the potential of rule-based adaptivity.

\cite{Fatahi} designed a personality-based adaptive system using MBTI, emotional cues, quizzes, and feedback. Within the sample of $n=222$ students, those who used the adaptive version showed reduced hint usage (29.94\%) and higher engagement on average.

These studies demonstrate the potential positive impact of intelligent learning systems, especially towards personalised educational pathways enabled by \gls{ai}. However, challenges remain in defining clear adaptive targets, improving system usability, and validating effectiveness through broader real-world testing.

\subsection{Assessment and Evaluation in Adaptive Learning Systems}

Assessment and evaluation are key considerations in the design of adaptive learning systems, enabling continuous monitoring of learner progress and supporting real-time instructional adjustments. Unlike traditional models that rely on pre-defined summative tests, adaptive systems use ongoing, often AI-driven assessments to personalise learning and, if required, alert educators to the need for intervention. 

Automated quiz generation is a common method for collecting immediate performance data. For example, \cite{Trajkovski} developed an AI-based quiz generator that adjusts the difficulty of questions according to the student's progress. The findings showed that the approach improved engagement and retention. Such automated assessments offer scalable, low-latency feedback essential for personalised learning. 

Adaptive systems are increasingly incorporating affective and behavioural indicators. \cite{Fatahi} integrated personality traits and emotion detection to support more nuanced content adaptation, improving engagement despite limited participation. However, many studies lack standardised benchmarks or explicit learning targets, making it difficult to measure pedagogical impact consistently \citep{Martin}.

Feedback remains a critical component of adaptive learning. \cite{Kolekar} showed that timely system-generated feedback helps learners quickly identify and correct misunderstandings. However, evaluations often focus on technical performance, such as algorithm accuracy or system responsiveness, rather than deeper educational outcomes \citep{Kabudi}. This results in limited insight into long-term learning and skill development.

Key challenges include scalability, validity of assessments, data privacy, and ethical concerns. As \cite{Chiu} notes, evaluation frameworks must go beyond test scores to consider motivation, emotional well-being, and autonomy. A more holistic and multidimensional approach is necessary to understand the educational impact of adaptive learning technologies fully.

\subsection{Multilingualism in Education}

Multilingualism in education has been widely studied, particularly in linguistically diverse contexts such as Nigeria, which comprises over 250 ethnic groups and more than 500 languages \citep{kori}. Research consistently shows that instruction in learners’ first languages supports comprehension, cognitive development, and academic achievement \citep{cummins, benson}. Reflecting this evidence, {\href{https://www.unesco.org/en/articles/why-mother-language-based-education-essential}{UNESCO}} advocates mother-tongue-based multilingual education (MTB-MLE), particularly in the early years of schooling.

Within the Nigerian context, \cite{obiakor} notes that although the National Policy on Education recommends the use of indigenous languages at the primary level, implementation remains inconsistent. According to \cite{bamgboṣe} and \cite{adegbija}, the continued dominance of English in classrooms despite being a second language for most learners has been associated with reduced classroom engagement and poorer learning outcomes. These scholars further contend that such practices reinforce linguistic hierarchies and marginalise local languages.

Concerns regarding the feasibility of multilingual education are highlighted by \cite{smits}, who identifies logistical complexity and resource limitations as major barriers. In contrast, \cite{heugh} demonstrates through studies in Ethiopia and South Africa that multilingual education can be successfully implemented through phased planning and sustained policy support. More recently, \cite{Bubou} argues that digital technologies, including localised and open-access educational platforms, offer viable mechanisms for supporting multilingual instruction, provided they are integrated within appropriate pedagogical and institutional frameworks.

\subsection{Challenges to Implementing Adaptive Learning Technologies in Nigeria}

Infrastructural deficits, inadequate funding, and low digital literacy constrain the adoption of adaptive learning technologies in Nigeria’s higher education sector. Despite their potential to support personalised learning, the national e-learning framework remains fragmented and underdeveloped.

\cite{Eli} reports that before COVID-19, the \gls{noun} was the only institution with a functional Open and Distance Learning (ODL) system, which was insufficient as a national model. Key barriers include unreliable electricity, poor internet access, and education budgets ranging from 4 to 7.24\%, far below UNESCO’s recommended 15 to 20\%. The study emphasises the importance of collaboration between the NUC and institutions in developing a cohesive e-learning curriculum.

\cite{Bubou} identify further obstacles such as low teacher capacity, digital illiteracy, technophobia, inadequate ICT infrastructure, high implementation costs, and the exclusion of visually impaired learners due to inaccessible content. The lack of inclusive design research also limits equitable access.

Similarly, \cite{Abaa} highlight challenges at \gls{noun}, including poor infrastructure, limited culturally relevant content, and low digital literacy. They recommend strengthening technological infrastructure, improving teacher training, ensuring data privacy, and producing high-quality local content.

Overall, the literature indicates that Nigeria faces interconnected challenges involving infrastructure, funding, and digital skills that hinder large-scale adoption of adaptive learning. While initiatives like \gls{noun} provide valuable insights, a coordinated national strategy focused on reliable infrastructure, inclusive design, local content development, and capacity building is essential for the effective and equitable implementation of these initiatives.

\section{Research purpose and research questions}\label{sec3}

The primary aim of this research is to enhance educational access and learning outcomes by integrating smart technologies and \gls{ai} to develop AI-enhanced, personalised, and adaptive learning tools. By addressing limitations in current educational systems and promoting curriculum internationalisation, the study seeks to provide learner-centred experiences that are inclusive, adaptive, and capable of improving both the quality and reach of education. To achieve this objective, the following research questions are formulated:

\textbf{RQ1:} How can AI be effectively leveraged in personalised and adaptive learning systems to support linguistic and learning diversity in low-resource environments?

\textbf{RQ2:} In what ways can technology help bridge disparities in educational delivery and ensure equitable standards of education across regions?

\textbf{RQ3:} How can AI function as an assessor and provide adaptive feedback in under-resourced educational systems?

\textbf{RQ4:} How can students be holistically assessed beyond academic achievement to reflect broader aspects of learning and development?

\section{Proposed Methodology}\label{sec4}
This paper proposes an AI-enhanced personalised and adaptive learning methodology designed to address linguistic, pedagogical, and infrastructural challenges in under-resourced educational contexts. 

\begin{figure*}[htb!]
    \centering
    \includegraphics[width=\linewidth, height=10cm, keepaspectratio]{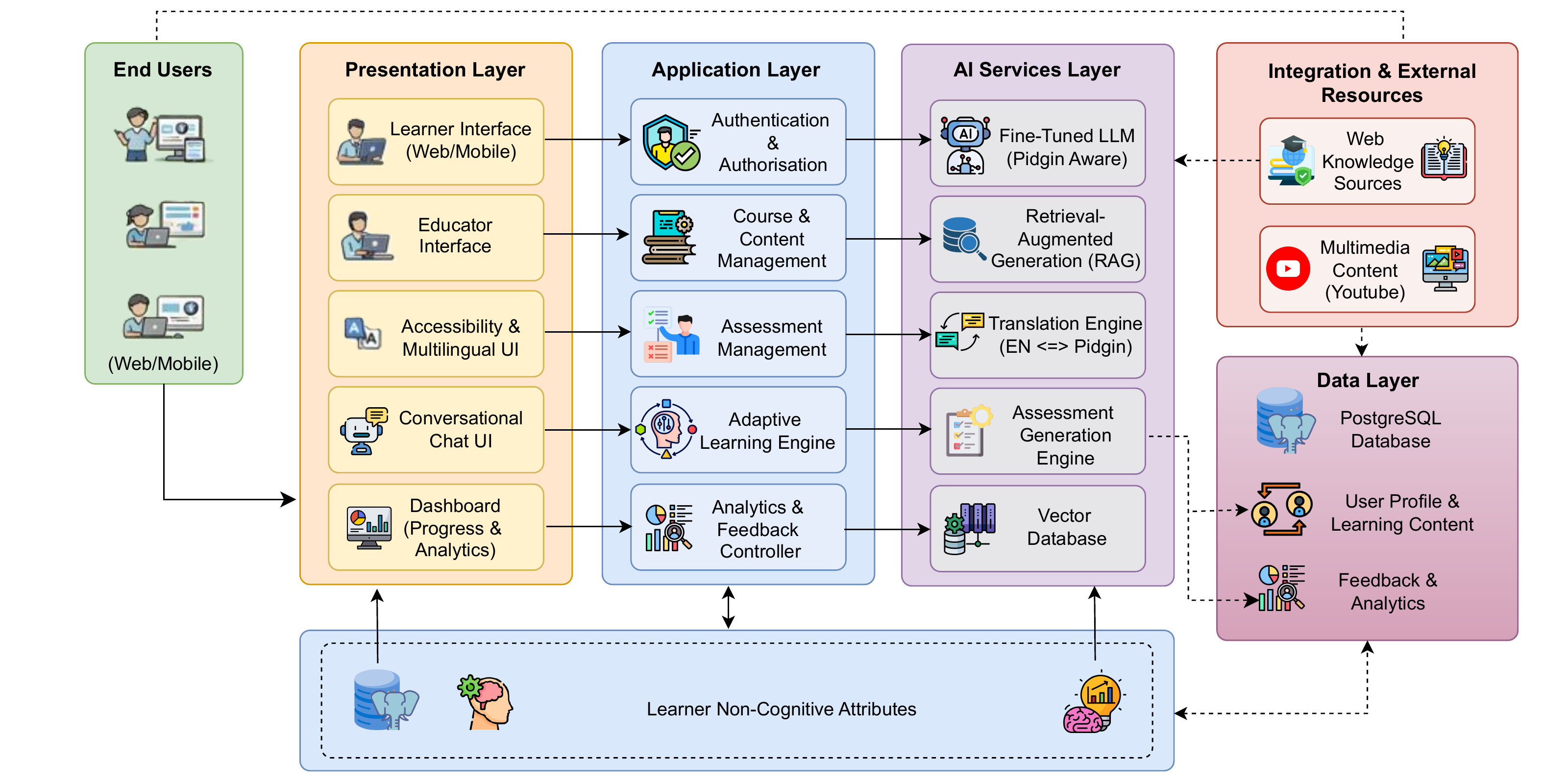}
    \caption{Conceptual architecture of the AI-enhanced personalised and adaptive learning platform. The platform adopts a modular, layered architecture comprising presentation, application, AI services, data, and integration layers. Fine-tuned large language models and retrieval-augmented generation support multilingual, context-aware learning, while a rule-based adaptive assessment framework integrates cognitive and non-cognitive learner attributes to enable personalised feedback, learning adaptation, and inclusive educational delivery.}
    \label{fig:exp1}
\end{figure*}
 
The methodology is operationalised through a modular, layered system architecture as depicted in Figure.~\ref{fig:exp1}, integrating conversational AI, adaptive assessment, and multilingual content delivery to support inclusive and context-aware learning.
The methodology begins with a learner-centred interaction layer, where students and educators access the platform through web or mobile interfaces. User inputs, learning activities, and assessment interactions are processed within the application layer, which orchestrates authentication, course management, assessment logics, and adaptive decision-making. A key component of the methodology is the adaptive learning engine, which personalises learning pathways by combining cognitive indicators in the form of assessment scores with selected non-cognitive attributes such as age, level of experience, motivation, language proficiency, and available study time.

To deliver intelligent and context-aware learning support, the methodology integrates an AI services layer comprising fine-tuned \gls{llm}, \gls{rag}, and a bilingual translation engine. \gls{rag} enhances response accuracy by grounding \gls{llm} outputs in semantically retrieved learning resources stored in a vector database, while fine-tuning ensures linguistic alignment with Nigerian Pidgin English. The translation workflow enables seamless switching between English and Pidgin, improving accessibility and learner engagement.

Persistent learner data, content, and analytics are managed within the data layer, supporting continuous feedback and performance monitoring. External knowledge sources and multimedia content are incorporated when internal retrieval confidence is insufficient, ensuring content relevance and adaptability.

\section{System Design and Technical Architecture}\label{sec5}
\subsection{System Design}

\begin{figure}
    \centering
    \includegraphics[width=\linewidth]{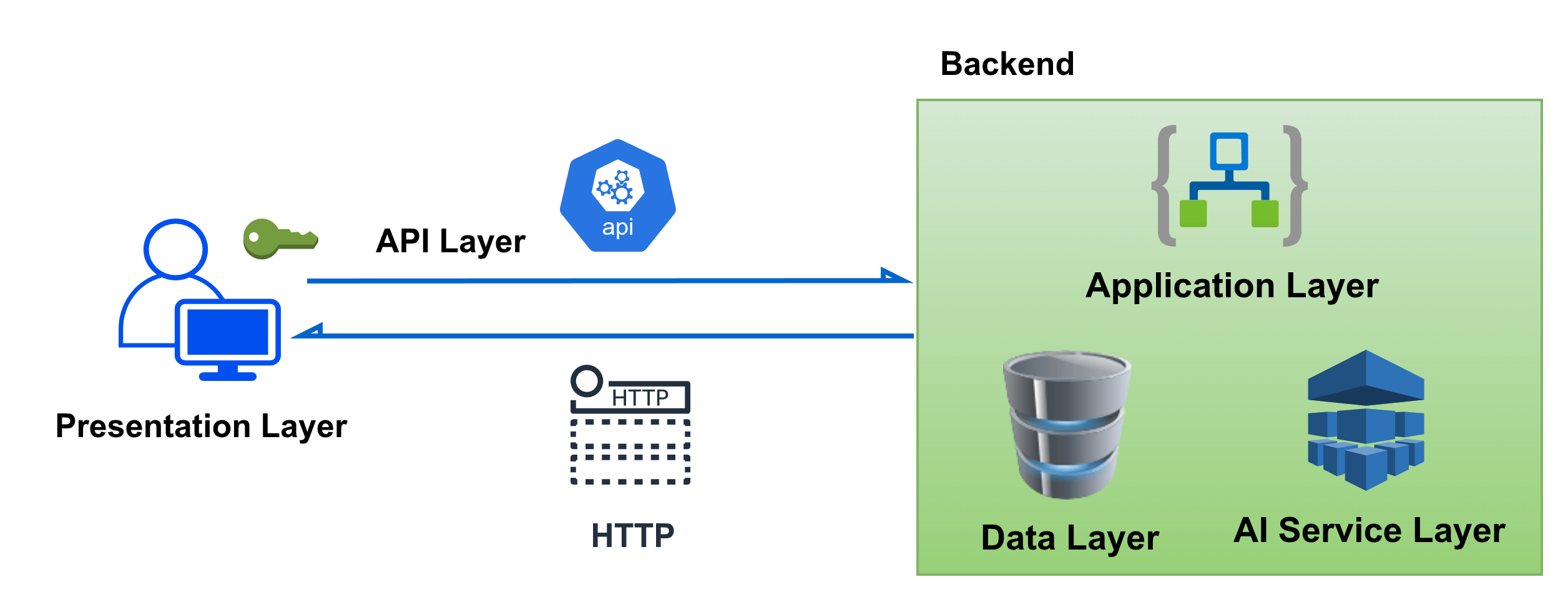}
    \caption{High-Level System Architecture Diagram.}
    \label{fig:sys_design}
\end{figure}

The system design, illustrated in Figure.~\ref{fig:sys_design}, adopts a modular and layered architecture for the AI-enhanced personalised and adaptive learning platform. The design emphasises inclusivity, personalisation, adaptability, and usability for diverse learner groups, while supporting scalability, maintainability, and seamless integration of AI-driven components. The architectural choices align with the functional requirements identified during requirements analysis.

The platform is structured into four main layers. The \textit{Presentation Layer}, implemented using React and Vite, manages user interaction and provides a lightweight and accessible interface, incorporating features such as text magnification to support inclusive access. The \textit{Application Layer}, developed with FastAPI, handles the core system logic, including user authentication, course management, assessment management, adaptive evaluation, conversational AI functionality, and performance analytics.

The \textit{Data Layer}, powered by PostgreSQL, is responsible for storing user profiles, learning content, assessment records, grades, and AI-generated feedback, ensuring data reliability and efficient retrieval. Finally, the \textit{AI Services Layer} hosts the transformer-based language model and \gls{rag} components, delivering personalised and context-aware responses through secure backend APIs.

\subsection{Conversational AI}

\begin{figure}
    \centering
    \includegraphics[width=1\linewidth]{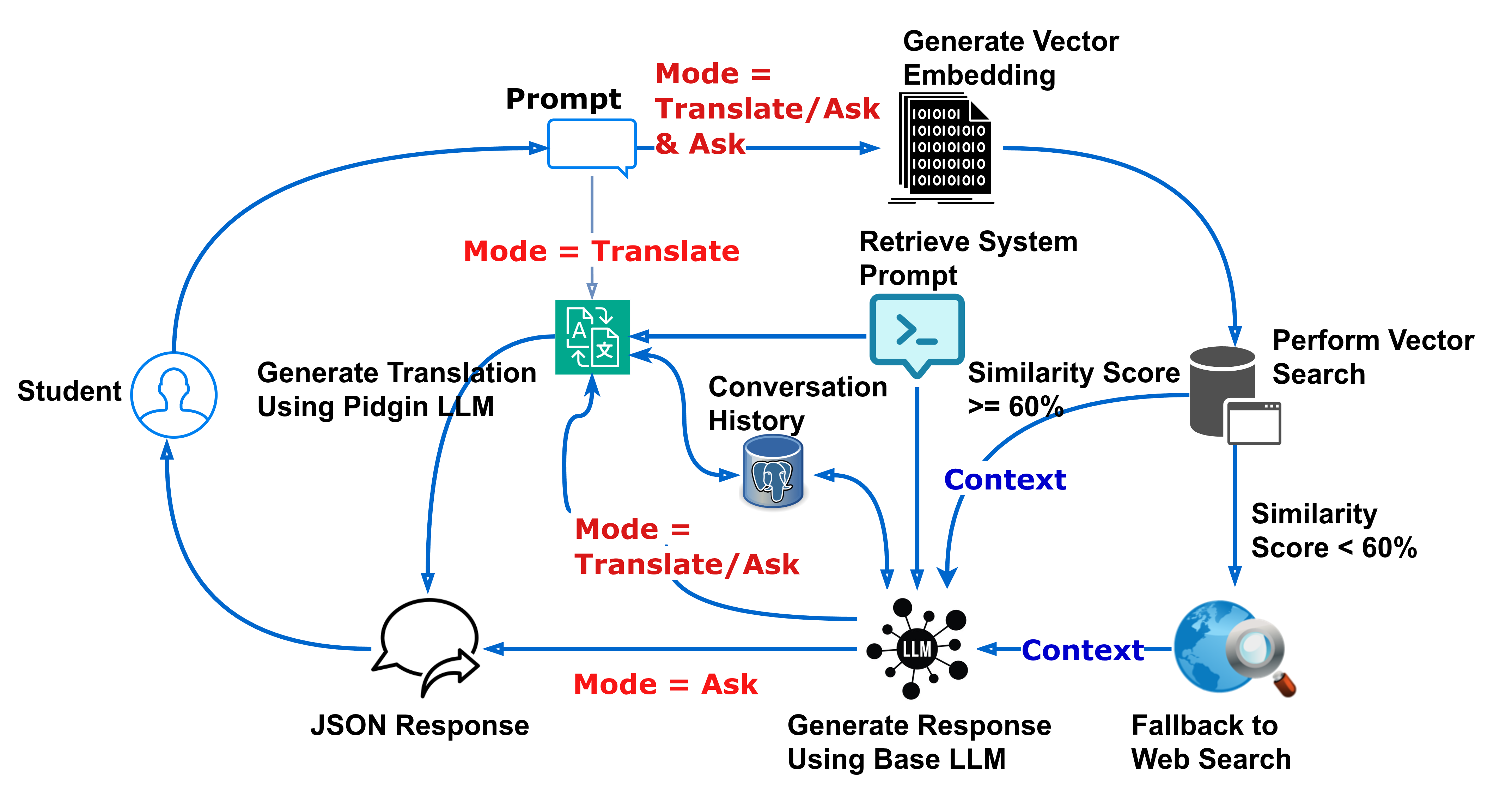}
    \caption{Conversation AI Workflow.}
    \label{fig:conver_ai}
\end{figure}

The conversational AI workflow, illustrated in Fig.~\ref{fig:conver_ai}, is powered by a combination of \gls{llm} and \gls{rag} and is implemented as a chatbot to support students through adaptive and context-aware interactions. The system operates in three modes: \textit{Translate}, \textit{Ask}, and \textit{Ask/Translate}.

In \textit{Translate} mode, the system employs a fine-tuned \gls{llm} guided by a predefined system prompt to translate the user prompt into Pidgin English. Conversation history is incorporated to maintain contextual coherence across interactions.

In \textit{Ask} mode, \gls{rag} is utilised to enhance response relevance. The user’s query is embedded and matched against a vector database, with web-based retrieval triggered when the similarity score falls below a 60\% threshold. Retrieved documents, together with the conversation history and system prompt, are passed to the \gls{llm} to generate a contextually grounded response in English.

The \textit{Ask/Translate} mode integrates both processes by first generating an enriched English response using the \textit{Ask} mode and subsequently translating the output into Pidgin English using the \textit{Translate} mode. This design enables accurate information retrieval while ensuring accessibility for users who speak Pidgin English.


\subsection{AI-driven assessment and adaptive feedback assistants.}

Learner performance within the system is evaluated using an AI-supported, rule-based assessment framework that integrates both cognitive and non-cognitive factors to enable personalised and adaptive learning. Academic grades are derived from assessments automatically generated and graded by an AI model using the course materials, while selected non-cognitive attributes are incorporated to contextualise learner performance. This hybrid approach supports informed educational decision-making while maintaining transparency and interpretability.

The rule-based framework is informed by established principles from pedagogy and cognitive science, ensuring consistency, fairness, and explainability in performance evaluation. By explicitly defining the contribution of each factor, the system avoids the opacity often associated with purely data-driven models, making it suitable for educational environments where accountability is essential.

The non-cognitive attributes considered include age, health status, learner motivation, available study time, language proficiency, and prior learning experience. These attributes provide a more holistic representation of learner readiness and contextual learning constraints, enabling the system to tailor feedback and learning pathways more effectively.

Let \(A \in [0,100]\) denote the academic score and \(N_1, \ldots, N_k \in [1,5]\) the non-cognitive scores.  
The average non-cognitive score is:
\begin{equation}
N = \frac{1}{k} \sum_{i=1}^{k} N_i
\end{equation}

\noindent The non-cognitive adjustment factor is defined as:

\begin{equation}
\mathrm{NC} = \frac{5 - N}{5}
\end{equation}

\noindent with a maximum non-cognitive weight \(w_N = 0.1\), the final performance score is computed as:
\begin{equation}
\mathrm{Performance} = A + \left( \frac{N}{5} \times 100 \times w_N \times \mathrm{NC} \right)
\label{eq:performance}
\end{equation}

\noindent The resulting performance score informs adaptive learning decisions, including personalised content delivery and targeted feedback interventions. Although machine learning-based approaches could be applied, this study adopts a rule-based method due to its clarity, interpretability, and suitability for educational decision-making.

\subsection{Course Management and Engagement}

\begin{figure}
    \centering
    \includegraphics[width=1.05\linewidth]{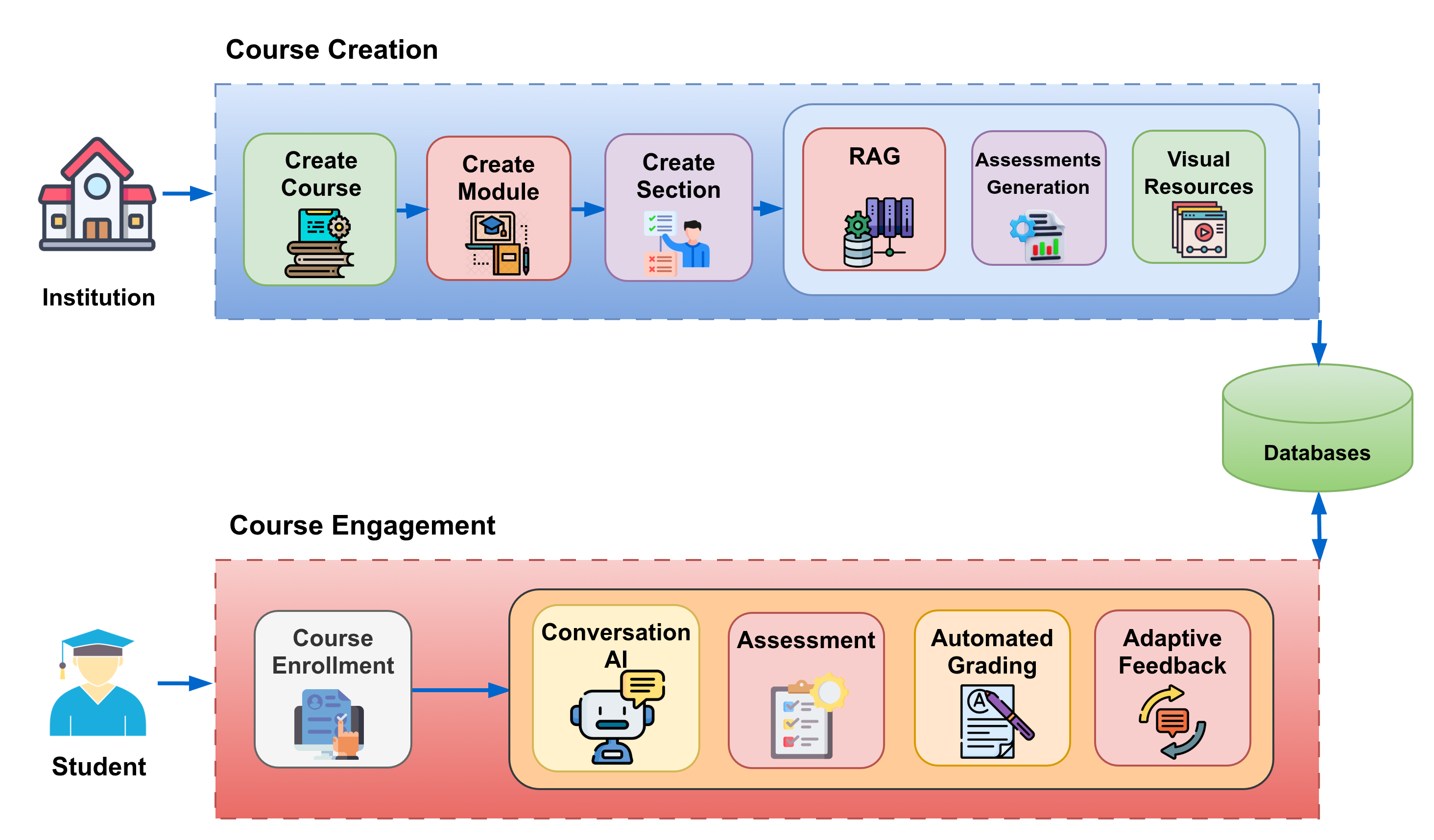}
    \caption{Course Management and Engagement Workflow.}
    \label{fig:course_man}
\end{figure}

The workflow illustrated in Figure.~\ref{fig:course_man} presents a high-level overview of course creation and learner engagement within the AI-driven platform. Educational institutions initiate the process by creating courses and organising them into modules and sections. An internal content-processing pipeline then refines instructional materials using \gls{rag}, generates assessments via the base \gls{llm}, and retrieves relevant YouTube videos to support multimodal learning.

From the learner's perspective, the workflow begins with course enrolment and progresses through lesson interaction, chatbot-assisted learning, assessment completion, and the receipt of real-time grading and adaptive feedback. This integrated process delivers a personalised and responsive learning experience, supported by a robust data infrastructure built on PostgreSQL and Qdrant for efficient content retrieval and learner analytics.

\section{Experimental Methodology}\label{sec6}

This research adopts a structured methodological framework to guide the design, development, deployment, and evaluation of an AI-enhanced personalised and adaptive educational system. The methodology integrates a fine-tuned \gls{llm} with interactive system development, emphasising reproducibility, scalability, and applicability in multilingual and under-resourced contexts. An iterative approach was used, allowing for continuous refinement of model training, system architecture, and user-centred design in alignment with the research objectives.

\subsection{Data Collection}


Textual data in Nigerian Pidgin English were collected to support the fine-tuning and evaluation of language models for multilingual and low-resource contexts. The dataset was web-scraped from publicly accessible news platforms, including \href{https://www.bbc.com/pidgin}{BBC Pidgin} and \href{https://pidgin.prime9ja.com.ng}{Prime9ja}, where Nigerian Pidgin is routinely used for informal reporting. These sources were selected to capture a broad range of contemporary linguistic usage, encompassing narrative, conversational, and domain-specific expressions.

Data acquisition was performed using automated web scraping techniques, as illustrated in Algorithm~\ref{alg:data_collection}. 
To ensure responsible data collection, scraping was limited to publicly available content, excluded user-restricted pages, and incorporated request rate-limiting to minimise server load. Collected data were handled in accordance with ethical research guidelines.


The raw corpus was subsequently cleaned and normalised through a series of preprocessing steps, including lowercase conversion, removal of empty and duplicate lines, filtering of non-textual elements, and basic text normalisation. Following these cleaning procedures, the final dataset comprises 416,343 high-quality Nigerian Pidgin English text entries spanning news reporting, commentary, and informal discourse. A representative sample of the dataset is shown in Figure~\ref{fig:sample_data}, highlighting the linguistic diversity of Nigerian Pidgin, including local idioms, non-standard grammar, code-mixed expressions, and domain-specific terminology. This diversity is essential for training and evaluating language models designed to operate effectively in multilingual, real-world educational settings.

\begin{algorithm}[H]
\caption{Data Collection of Nigerian Pidgin English.}
\label{alg:data_collection}
\begin{algorithmic}[1]
\State Initialise empty text corpus $D$
\For{each source category}
    \For{each page index in the category}
        \State Fetch HTML content of the page
        \State Extract article links from the page
        \For{each article URL}
            \State Fetch HTML content of the article
            \State Extract paragraph text segments
            \For{each paragraph segment}
                \State Normalise text: lowercasing, whitespace trimming etc.
                \State Append normalised text to corpus $D$
            \EndFor
        \EndFor
        \State Pause for a fixed interval to enforce rate limiting
    \EndFor
\EndFor
\State Save corpus $D$ to persistent storage
\end{algorithmic}
\end{algorithm}







\begin{figure}
    \centering
    \fbox{\parbox{\linewidth}{
pipo for the procession just dey stand dey watch am as e dey scream say im body dey burn.\\
later e go university of ibadan wey e get b.sc. for biochemistry.\\
1638 - di english again bin attempt to settle.\\
na justice simon aboki bin preside ontop di case for di hearing.\\
strikers: princess d. marfo, sherrifatu sumaila, alice kusi, abigail kofi kim\\
na im be di current secretary to di goment of di federation.\\
group c: the gambia vs cameroon, bouake (17:00)\\
60' freekick for nigeria against sao tome and principe.\\
di pipo wey visit am na youths from di lga no be thugs\
``anything wey happun dia i no dey dia, so make im tell police di truth''.\\
``di pipo wey visit am na youths from di lga no be thugs''
    }}
    \caption{Representative sample of the Nigerian Pidgin English dataset, showing culturally and contextually rich expressions.}
    \label{fig:sample_data}
\end{figure}

\subsection{Data Preprocessing}

Data preprocessing is a crucial step in transforming raw, unstructured data into a structured format that is suitable for model training. In this study, the Pidgin English text collected through web scraping required extensive preprocessing due to informal grammar, variable spelling, and inconsistent punctuation. The preprocessing pipeline involved cleaning the text using regular expressions to remove noise, domain-specific artefacts, and redundant characters, followed by deduplication to ensure sample diversity. The cleaned data were then formatted into instruction-style conversational pairs for natural language processing tasks.

Subsequently, the dataset was converted into a Hugging Face–compatible format to enable seamless integration with the training pipeline and tokenised using the \texttt{AutoTokenizer} to transform text into numerical representations. This process resulted in a structured, deduplicated, and tokenised dataset, supporting effective model fine-tuning.

\subsection{Modelling}

The modelling approach adapts a pre-trained language model for educational interactions in Nigerian Pidgin English, ensuring alignment with the platform’s pedagogical objectives. Transformer-based architectures, such as GPT and LLaMA, are typically trained on high-resource languages and often underperform in low-resource contexts like Pidgin English educational dialogue. To address this, the model was fine-tuned on the collected Pidgin English dataset to enhance contextual understanding and response quality.

Parameter-efficient fine-tuning was implemented using LoRA, which freezes most of the base model parameters and updates only small, trainable low-rank adapter matrices within the transformer layers. This strategy substantially reduces memory requirements while enabling effective domain adaptation, making it feasible to fine-tune large language models on consumer-grade GPUs.

\subsection{Evaluation}

Model evaluation is essential for assessing both performance and generalisation, particularly in multilingual and low-resource settings where automated metrics alone may not fully capture output quality. In this study, evaluation was conducted along three complementary dimensions: \textit{computational and semantic evaluation}, \textit{cultural and linguistic evaluation}, and \textit{efficiency evaluation}. Computational and semantic metrics provide reproducible quantitative benchmarks; efficiency evaluation examines inference latency and computational cost; and cultural and linguistic evaluation captures pragmatic, contextual, and sociolinguistic characteristics of Nigerian Pidgin English that are often overlooked by automated measures.


The evaluation dataset comprised 14 sample prompts, selected to capture typical educational and conversational scenarios. The prompts consisted of common science questions. Each prompt was processed using the three fine-tuned models, with outputs from OpenAI gpt-4o serving as reference responses. These references provided a widely adopted baseline for comparative analysis rather than an absolute ground truth. The prompt-response pairs are publicly accessible\footnote{Prompt-response: \url{https://yellow-fania-55.tiiny.site/}}

\subsubsection{Computational and Semantic Evaluation}

Computational and semantic evaluation was conducted using BLEU, ROUGE-L, perplexity, Distinct-1, and BERT Score. Together, these metrics assess lexical overlap, structural alignment, semantic similarity, fluency, and lexical diversity, offering a balanced view of model performance across surface-level and meaning-oriented dimensions.

\textbf{BLEU} (Bilingual Evaluation Understudy), originally proposed by \cite{papineni2002bleu}, measures n-gram overlap between generated responses and reference texts. Despite known limitations in capturing semantic variation, BLEU remains a standard indicator of syntactic similarity. In this study, BLEU was computed with $N=3$ and uniform weights \(w_n = 1/N\):
\begin{equation}
\mathrm{BLEU} = \mathrm{BP} \, \exp \left( \sum_{n=1}^{N} w_n \log p_n \right)
\end{equation}

\textbf{ROUGE-L} \citep{lin2004rouge} evaluates similarity based on the longest common subsequence (LCS), capturing sentence-level structure and word order preservation:
\[
\text{ROUGE-L} = \frac{(1+\beta^2) \cdot \text{LCS}}{R + \beta^2 P}.
\]

\textbf{Perplexity} quantifies model confidence and fluency by measuring how well a language model predicts a sequence of tokens \citep{jelinek1977perplexity}:
\[
\text{Perplexity} = \exp\left(-\frac{1}{N}\sum_{i=1}^{N} \log p(w_i)\right)
\]

\textbf{Distinct-1} measures lexical diversity by computing the proportion of unique unigrams in generated text \citep{li2016diversity}, helping identify repetition and mode collapse:
\[
\text{Distinct-1} = \frac{|\text{unique unigrams}|}{|\text{total tokens}|}
\]

\textbf{BERTScore} \citep{zhang2019bertscore} evaluates semantic similarity using contextual embeddings from transformer-based models, enabling meaning-based comparison even when surface forms differ. This property is particularly important for Nigerian Pidgin English, which exhibits flexible grammar and expressive variation.

\subsubsection{Cultural and Linguistic Evaluation}
To complement the computational evaluation, a human-centred cultural and linguistic assessment was conducted. Recent scholarship on generative AI in education underscores the importance of trustworthiness, contextual alignment, and rigorous evaluation beyond automatic metrics \citep{AI_Future_of_Education_Review}, particularly when deploying systems in culturally diverse settings. Model-generated responses in Nigerian Pidgin English were collected using a Google Form and evaluated by bilingual English–Pidgin speakers. Responses were rated on a five-point Likert scale (1–5) across fluency, coherence, relevance, and cultural appropriateness. Qualitative feedback was also collected to capture nuanced linguistic phenomena, such as idiomatic usage, pragmatic correctness, and natural phrasing.

\textbf{Fluency} assesses how naturally the text reads, focusing on grammatical correctness, sentence structure, and smoothness of expression.  

\textbf{Coherence} evaluates whether the response is logically consistent, well-structured, and easy to follow, with ideas and sentences flowing in a meaningful sequence.  

\textbf{Relevance} measures the extent to which the response addresses the input prompt accurately and appropriately, capturing the intended meaning or task.  

\textbf{Appropriateness} examines the alignment of language, idioms, and expressions with Nigerian Pidgin English norms, including colloquial usage, pragmatics, and contextually correct phrasing.

\subsubsection{Efficiency Evaluation}

Efficiency evaluation focused on inference latency under a consistent experimental setup. All quantised models were deployed on Google Colab using an NVIDIA T4 GPU. The average inference time per prompt was recorded to enable a comparative analysis of runtime efficiency across different quantisation levels, reflecting practical deployment considerations in resource-constrained environments.

\section{Results and Discussion}\label{sec7}
This section presents and discusses the results of the experimental evaluation, situating both quantitative and qualitative findings within the objectives of the proposed AI-enhanced \gls{pal} platform. The full system implementation is publicly accessible\footnote{STEMHub: \url{https://www.stemhub-learning.com/}}. The fine-tuned model is available for download\footnote{Zenodo: \url{https://doi.org/10.5281/zenodo.18451305}, Hugging Face: \url{https://huggingface.co/Guavacoderepo/pidgin-llama}}, and the dataset used for training the model is also publicly available\footnote{Zenodo: \url{https://doi.org/10.5281/zenodo.18467116}, Hugging Face: \url{https://huggingface.co/datasets/Guavacoderepo/gclm-pidgin-text-corpus}}. The evaluation was designed to address the research questions outlined in Section~\ref{sec3}, namely: whether the platform can deliver semantically coherent and contextually appropriate learning interactions in multilingual, low-resource settings (RQ1); whether fine-tuning \gls{llm} on Nigerian Pidgin English improves linguistic and cultural alignment (RQ2); and how quantisation-induced efficiency trade-offs affect practical deployability (RQ3).

The analysis integrates automatic evaluation metrics assessing lexical alignment, fluency, semantic similarity, and lexical diversity with human-centred assessments of fluency, coherence, relevance, and cultural appropriateness in Nigerian Pidgin English. This multi-dimensional approach enables a holistic evaluation that extends beyond surface-level accuracy to capture pedagogical usefulness and contextual relevance.
Across the three fine-tuned and quantised models, several consistent patterns emerge. All models demonstrate adequate semantic preservation, as indicated by BERTScore and ROUGE-L, despite low BLEU scores that reflect the limitations of n-gram-based metrics in low-resource and non-standardised languages. Efficiency results reveal clear trade-offs between quantisation level and inference latency, with smaller models achieving faster response times while maintaining acceptable semantic quality. Human evaluations further confirm that fine-tuning on Nigerian Pidgin enhances both linguistic naturalness and cultural appropriateness, collectively demonstrating that the proposed platform strikes a balance between performance quality, efficiency, and inclusivity.

\subsection{Computational and Semantic Evaluation}
Table~\ref{tab:auto_metrics} summarises the performance of the quantised language models and offers insight into how quantisation affects semantic coherence, fluency, and structural fidelity. These results directly inform RQ1, which investigates the ability of the proposed AI-enhanced \gls{pal} platform to deliver semantically coherent learning interactions, as well as RQ3, which examines the efficiency–quality trade-offs relevant to deployment in resource-constrained educational settings. We evaluate the results across four dimensions:

\begin{table}[htbp]
\centering
\caption{Automatic Evaluation Metrics for Quantised Models.}
\label{tab:auto_metrics}
\small
\begin{tabular}{l c c c c c}
\toprule
\textbf{Model} & \textbf{BLEU} & \textbf{ROU-L} & \textbf{PPL $\downarrow$} & \textbf{Dist-1} & \textbf{BERT-S} \\
\midrule
LLM-Q4 & 0.0121 & 0.3005 & 61.50 & 0.0616 & 0.8516 \\
LLM-Q5 & 0.0120 & 0.2840 & 76.93 & 0.0711 & 0.8580 \\
LLM-Q8 & 0.0120 & 0.3120 & 78.86 & 0.0658 & 0.8590 \\
\bottomrule
\end{tabular}
\end{table}
\textbf{Lexical and Structural Alignment:} The results on the BLEU scores are uniformly low, indicating minimal exact n-gram overlap between generated and reference responses. The consistently low BLEU scores across all models indicate limited exact n-gram overlap between generated and reference responses. Rather than signalling poor performance, this behaviour reflects the expressive flexibility of Nigerian Pidgin English and the open-ended nature of educational dialogue, where semantically valid responses may differ substantially in surface form. In this context, BLEU provides limited insight into instructional quality and reinforces the need for complementary semantic metrics when evaluating low-resource and non-standardised languages.

However, higher-bit quantisation (LLM-Q8) achieved the highest ROUGE-L score (0.3120), suggesting that increased numerical precision better preserves structural coherence and sentence-level flow. ROUGE-L’s reliance on longest common subsequence alignment makes it more sensitive to discourse structure than n-gram overlap alone, and the observed improvement indicates that higher-bit representations retain syntactic and sequential patterns that are important for pedagogical clarity and stepwise explanation. This structural consistency is particularly important in adaptive learning contexts, where explanations must remain logically ordered and easy to follow.

\textbf{Fluency and Diversity:} Interestingly, LLM-Q4 recorded the lowest, indicating higher confidence in token prediction and stronger local fluency relative to the other quantised models. This suggests that lower-bit quantisation can preserve surface-level linguistic smoothness, even under constrained numerical precision. Lexical diversity, measured using Distinct-1, remains relatively stable across all models, with LLM-Q5 showing a marginal advantage in vocabulary richness (0.0711). The limited variation in diversity scores suggests that quantisation depth has minimal impact on the range of lexical expression in generated responses.

\textbf{Semantic Similarity:} LLM-Q8 achieves the highest BERTScore (0.8590), indicating stronger semantic alignment between generated outputs and reference responses. This suggests that higher bit-width quantisation better preserves contextual meaning and conceptual intent, which is particularly important for educational feedback tasks where semantic adequacy outweighs exact lexical matching.

\textbf{Synthesis:} The results demonstrate complementary strengths across quantisation levels. Lower-bit quantisation (Q4) is associated with improved fluency and computational efficiency, as reflected by lower perplexity and faster inference, while higher-bit models (Q8) provide stronger structural and semantic preservation, evidenced by higher ROUGE-L and BERTScore values. Rather than indicating a single optimal configuration, these findings highlight a trade-off between efficiency and semantic robustness. This trade-off directly addresses \textbf{RQ3}, demonstrating how quantisation choices influence the balance between deployability and pedagogical reliability in AI-driven personalised learning systems.

\subsection{Cultural and Linguistic Evaluation}
To complement the computational metrics, a human-centred cultural and linguistic evaluation was conducted to assess whether the proposed platform produces linguistically natural and culturally appropriate outputs in Nigerian Pidgin English, directly addressing \textbf{RQ1} and \textbf{RQ2}. A structured survey was distributed to 50 native Nigerian Pidgin speakers via Google Forms, yielding 18 valid responses. Participants evaluated model-generated outputs across fluency, coherence, relevance, and cultural appropriateness.

Listings~\ref{lst:pidgin_inference} and~\ref{lst:pidgin_prompt_response} illustrate the inference pipeline and a representative model response, respectively. The example output demonstrates clear sentence construction, appropriate use of Pidgin vocabulary, and contextually relevant explanations, reflecting the model’s ability to convey instructional content in an accessible and culturally grounded manner. Across responses, participants consistently rated outputs from the fine-tuned models as understandable and context-aware, confirming that fine-tuning on Nigerian Pidgin data improves linguistic alignment relative to non-adapted baselines.

\begin{lstlisting}[language=Python, caption={Inference using the fine-tuned Pidgin model}, label={lst:pidgin_inference}]
llm = Llama.from_pretrained(
    repo_id="Guavacoderepo/pidgin-llama",
    filename="llama-3.2-3b-instruct.Q4_K_M.gguf"
)

messages = [
    {"role": "system",
     "content": "You are a pidgin instructor. Explain clearly in less than 50 words."},
    {"role": "user",
     "content": "Explain artificial intelligence"}
]

llm.create_chat_completion(messages=messages)
\end{lstlisting}

\begin{lstlisting}[caption={Model response from the 4-bit quantised LLM-Q4 model}, label={lst:pidgin_prompt_response}]
{
  'id': 'chatcmpl-1f310250-315c-4b20-b016-8645f60ea353',
  'object': 'chat.completion',
  'created': 1767922707,
  'model': 'llama-3.2-3b-instruct.Q4_K_M.gguf',
  'choices': [{
    'index': 0,
    'message': {
      'role': 'assistant',
      'content': 'artificial intelligence na computer system wey fit do tins wey humans fit do, like think and learn. e dey use special software wey dey called algorithms to make decisions and solve problems. e fit also dey use machine learning to learn from data.'
    },
    'finish_reason': 'stop'
  }],
  'usage': {
    'prompt_tokens': 64,
    'completion_tokens': 55,
    'total_tokens': 119
  }
}
\end{lstlisting}


\subsubsection{Fluency}
\begin{figure}
\centering
\scriptsize
\begin{tikzpicture}
\begin{axis}[
    height=5.5cm,
    width=0.90\linewidth,
    ybar,
    bar width=18pt,
    symbolic x coords={LLM\_Q4, LLM\_Q5, LLM\_Q8, OpenAI gpt-4o},
    xtick=data,
    ylabel={Score},
    ymin=3.5,
    ymax=5,
    nodes near coords,
    axis background/.style={draw=black, thick},
]
\addplot coordinates {(LLM\_Q4,3.78) (LLM\_Q5,3.89) (LLM\_Q8,3.94) (OpenAI gpt-4o,4.28)};
\end{axis}
\end{tikzpicture}
\caption{Human evaluation results for fluency.}
\label{fig:fluency_eval}
\end{figure}
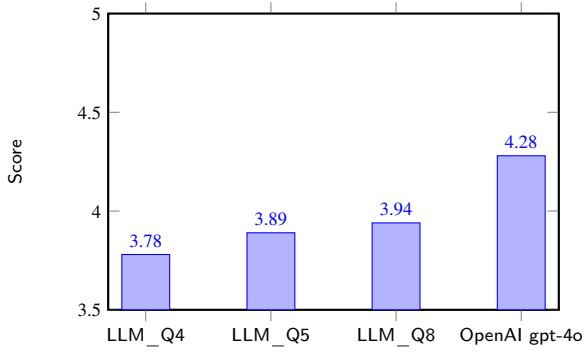

Fluency ratings reveal clear qualitative differences across quantisation levels. LLM-Q4 was generally intelligible but frequently exhibited repetition and code-mixing with standard English and Nigerian Pidgin, resulting in a less natural conversational flow. LLM-Q5 showed improved sentence structure and smoother phrasing, though traces of English-dominant vocabulary persisted. LLM-Q8 produced the most fluent and conversational outputs among the quantised models, closely aligning with the rhythm and informal register of natural Nigerian Pidgin. As illustrated in Figure~\ref{fig:fluency_eval}, the OpenAI gpt-4o reference model achieved the highest fluency overall, serving as an upper benchmark for human-like expression.

These reinforce the computational results by demonstrating that higher-bit quantisation and targeted fine-tuning enhance linguistic naturalness and cultural appropriateness. In doing so, the human evaluation provides qualitative evidence that the proposed AI-enhanced PAL platform can support meaningful and contextually appropriate learning interactions in multilingual settings, directly addressing \textbf{RQ1} and \textbf{RQ2}.

\subsubsection{Coherence}
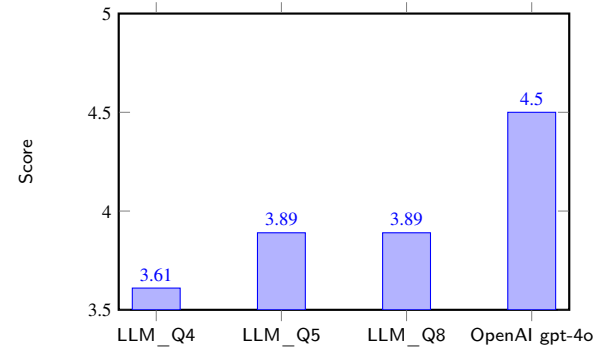
\begin{figure}
\centering
\scriptsize
\begin{tikzpicture}
\begin{axis}[
    height=5.5cm,
    width=0.90\linewidth,
    ybar,
    bar width=18pt,
    symbolic x coords={LLM\_Q4, LLM\_Q5, LLM\_Q8, OpenAI gpt-4o},
    xtick=data,
    ylabel={Score},
    ymin=3.5,
    ymax=5,
    nodes near coords,
    axis background/.style={draw=black, thick},
]
\addplot coordinates {(LLM\_Q4,3.61) (LLM\_Q5,3.89) (LLM\_Q8,3.89) (OpenAI gpt-4o,4.50)};
\end{axis}
\end{tikzpicture}
\caption{Human evaluation results for coherence.}
\label{fig:coherence_eval}
\end{figure}

LLM\_Q4 sometimes produced fragmented or disjointed responses that were difficult to follow. LLM\_Q5 improved in maintaining logical progression and linking ideas, resulting in clearer and more structured responses. LLM\_Q8 further enhanced discourse consistency, producing responses that were stable and easy to interpret. Figure~\ref{fig:coherence_eval} shows that the OpenAI gpt-4o reference model outperformed all quantised models, providing highly coherent responses, with well-connected ideas and smooth narrative flow suitable for professional and academic settings.

\subsubsection{Relevance}
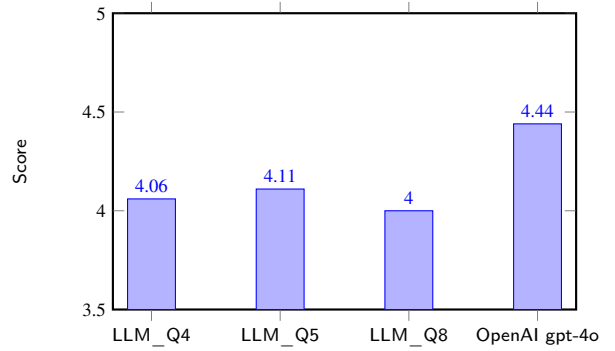
\begin{figure}
\centering
\scriptsize
\begin{tikzpicture}
\begin{axis}[
    height=5.5cm,
    width=0.92\linewidth,
    ybar,
    bar width=18pt,
    symbolic x coords={LLM\_Q4, LLM\_Q5, LLM\_Q8, OpenAI gpt-4o},
    xtick=data,
    ylabel={Score},
    ymin=3.5,
    ymax=5,
    nodes near coords,
    axis background/.style={draw=black, thick},
]
\addplot coordinates {(LLM\_Q4,4.06) (LLM\_Q5,4.11) (LLM\_Q8,4.00) (OpenAI gpt-4o,4.44)};
\end{axis}
\end{tikzpicture}
\caption{Human evaluation results for relevance.}
\label{fig:relevance_eval}
\end{figure}

LLM\_Q4 often answered prompts partially, occasionally missing key details. LLM\_Q5 improved prompt coverage and delivered more context-aware answers. LLM\_Q8 showed the most consistent relevance among the quantised models, closely aligning with the intent of the prompts. Figure~\ref{fig:relevance_eval} illustrates that the OpenAI gpt-4o reference model consistently produced highly relevant and focused responses, precisely capturing the prompts’ meaning while remaining concise and clear.

\subsubsection{Cultural Appropriateness}
\begin{figure}
\centering
\scriptsize
\begin{tikzpicture}
\begin{axis}[
    height=5.5cm,
    width=0.95\linewidth,
    ybar,
    bar width=18pt,
    symbolic x coords={LLM\_Q4, LLM\_Q5, LLM\_Q8, OpenAI gpt-4o},
    xtick=data,
    ylabel={Score},
    ymin=3.5,
    ymax=5,
    nodes near coords,
    axis background/.style={draw=black, thick},
]
\addplot coordinates {(LLM\_Q4,3.72) (LLM\_Q5,3.83) (LLM\_Q8,4.06) (OpenAI gpt-4o,4.22)};
\end{axis}
\end{tikzpicture}
\caption{Human evaluation results for cultural appropriateness.}
\label{fig:appropriateness_eval}
\end{figure}
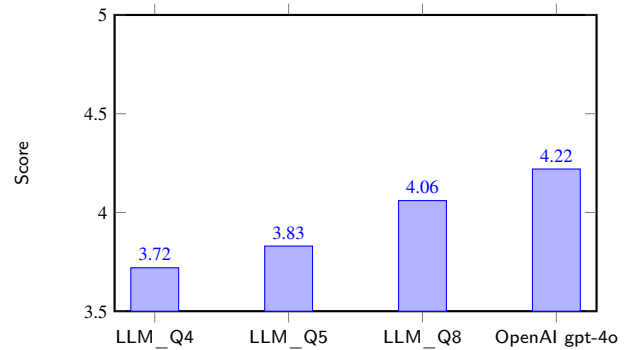

LLM\_Q4 frequently sounded like a direct English translation, missing colloquial nuances. LLM\_Q5 improved slightly, though English-heavy phrasing remained. LLM\_Q8 offered the best cultural alignment among the quantised models, with more natural expressions, yet some idiomatic richness was still lacking. As shown in Figure~\ref{fig:appropriateness_eval}, the OpenAI gpt-4o model demonstrated the strongest cultural grounding, using accurate Pidgin expressions and contextually appropriate phrasing that reflected local linguistic norms.

\subsubsection{Efficiency Evaluation}

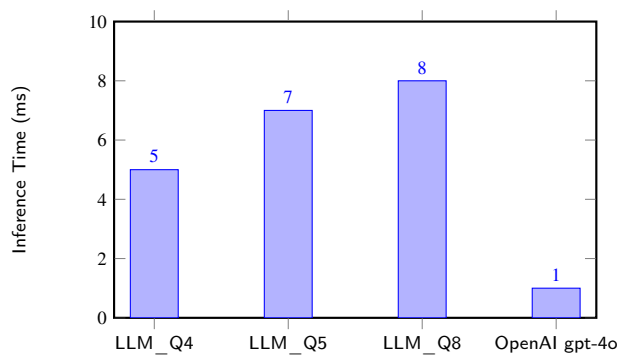
\begin{figure}
\centering
\scriptsize
\begin{tikzpicture}
\begin{axis}[
    height=5.5cm,
    width=0.95\linewidth,
    ybar,
    bar width=18pt,       
    symbolic x coords={LLM\_Q4, LLM\_Q5, LLM\_Q8, OpenAI gpt-4o},
    xtick=data,
    ylabel={Inference Time (ms)},
    ymin=0,
    ymax=10,
    nodes near coords,
    axis background/.style={draw=black, thick},
]

\addplot+ coordinates {(LLM\_Q4,5) (LLM\_Q5,7) (LLM\_Q8,8) (OpenAI gpt-4o,1)};

\end{axis}
\end{tikzpicture}

\caption{Average inference time of the three quantised language models.}
\label{fig:inference_time}
\end{figure}

Figure~\ref{fig:inference_time} illustrates the average inference time across models. The 4-bit, 5-bit, and 8-bit quantised models required 5~s, 7~s, and 8~s per inference, respectively, compared to 1~s for the OpenAI gpt-4o base model. These results highlight a clear latency increase with higher-bit quantisation, indicating a trade-off between computational efficiency and model configuration.

\begin{figure*}
    \centering
    \includegraphics[width=1\linewidth]{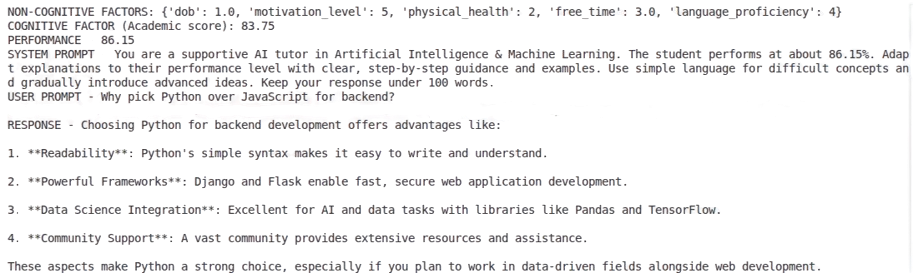}
    \caption{Performance evaluation and adaptive AI response for Student~1, where strong non-cognitive contributions combined with a high academic score result in minimal instructional scaffolding.}
    \label{fig:student1}
\end{figure*}

\begin{figure*}
    \centering
    \includegraphics[width=1\linewidth]{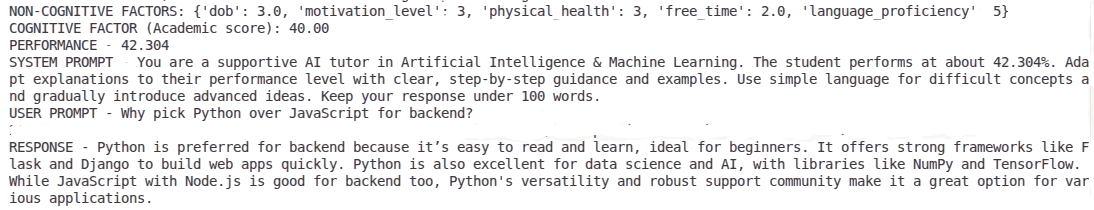}
    \caption{Performance evaluation and adaptive AI response for Student~2, illustrating how moderate non-cognitive contributions and lower academic performance lead to increased instructional support.}
    \label{fig:student2}
\end{figure*}

\subsection{Interactive Application and Model Integration}

Table~\ref{tab:functional_discussion} summarises the functional evaluation of the implemented AI-enhanced learning platform. The results highlight the system’s successful integration of key components, including user authentication, course and content management, \gls{rag}, \gls{llm}, and learner analytics. The platform supports reliable navigation, context-aware assessment generation, chatbot interactions, and performance-based feedback, with learner dashboards facilitating progress tracking and updates to non-cognitive data.

\begin{figure*}[t]
    \centering
    \includegraphics[width=0.9\linewidth]{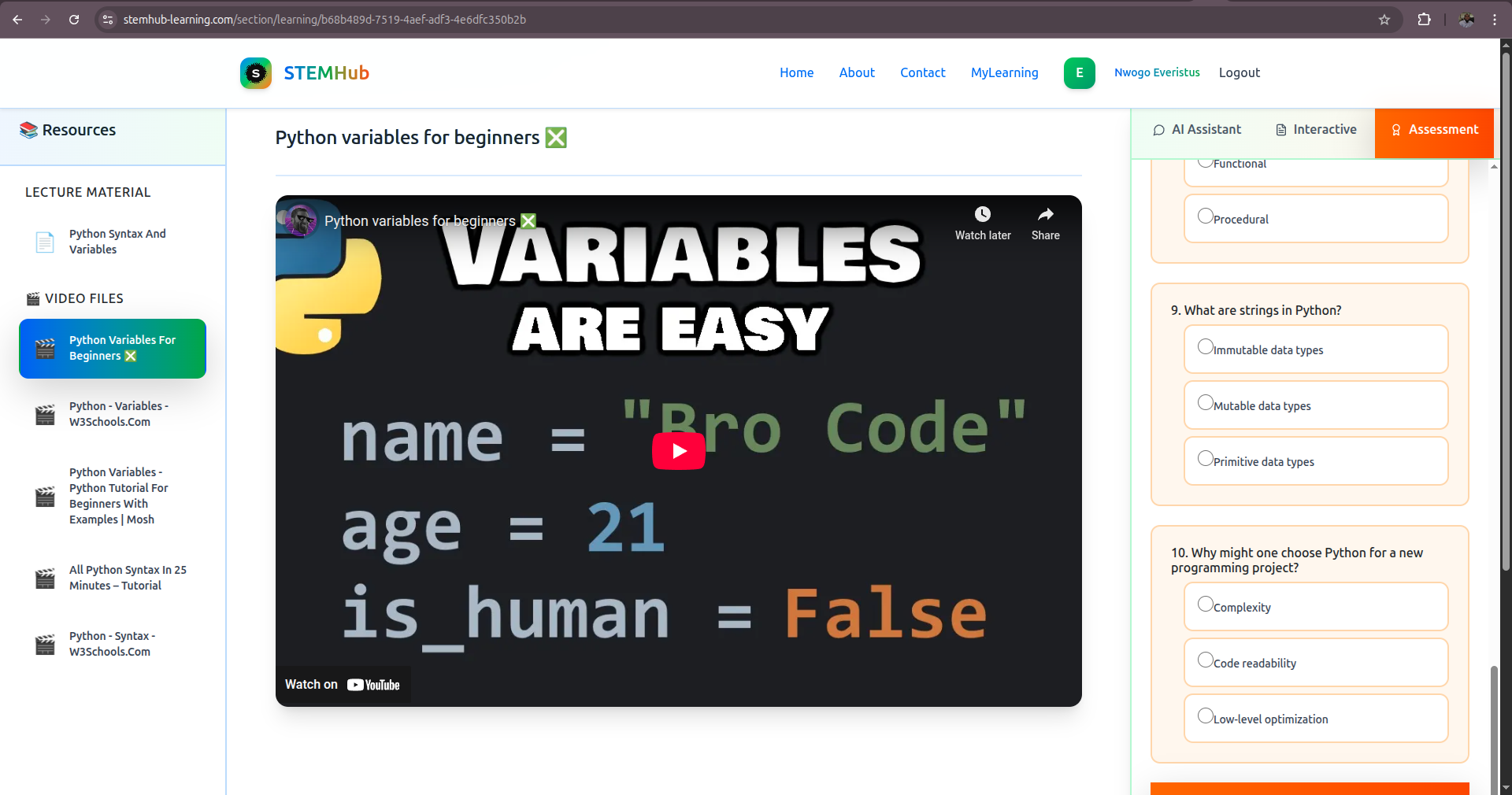}
    \caption{The STEMHub platform: student learning environment integrating core instructional content, external multimedia resources, and AI-generated adaptive assessments to support personalised and interactive learning.}
    \label{fig:learning_environment}
\end{figure*}

\begin{figure*}[t]
    \centering
    \includegraphics[width=0.9\linewidth]{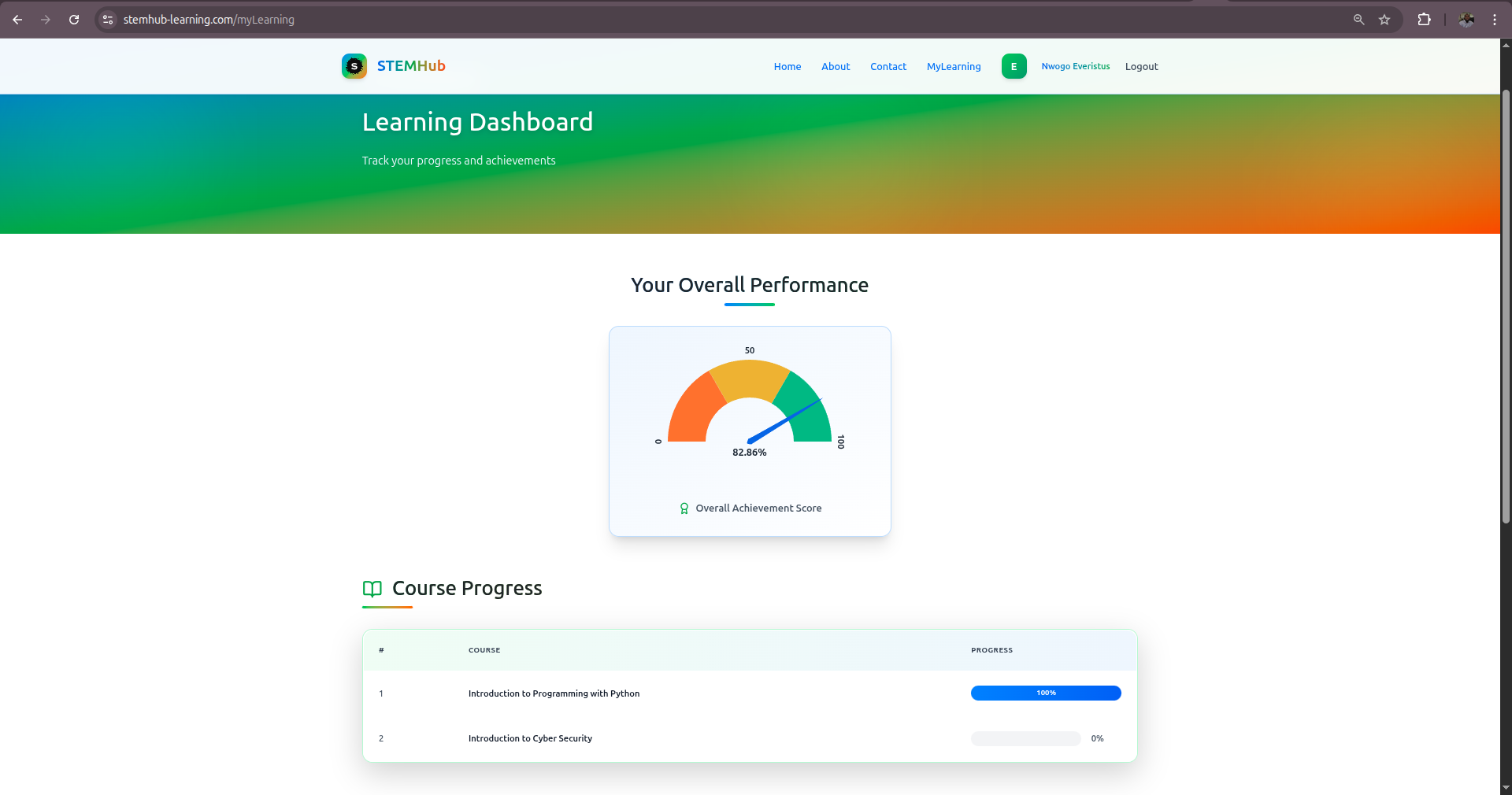}
    \caption{The STEMHub platform: learner dashboard displaying progress tracking, performance analytics, and personalised feedback to support adaptive learning and holistic learner modelling.}
    \label{fig:student_dashboard}
\end{figure*}

Figures~\ref{fig:learning_environment} and \ref{fig:student_dashboard} illustrate the platform in action, showing the integrated instructional environment with external resources and AI-generated assessments. The interface enables seamless navigation between course modules, interactive learning materials, and real-time feedback. The learner dashboard provides detailed tracking of academic performance, allowing students to visualise their progress over time. Additionally, the system supports personalised and adaptive responses, tailoring the learning experience to each student’s strengths and areas for improvement.

Figures~\ref{fig:student1} and~\ref{fig:student2} present two student cases interacting with the conversational learning system. Student~1 achieved a high academic score of 83.75 with favourable non-cognitive factors, resulting in an overall performance score of 86.15. In contrast, Student~2 obtained a lower academic score of 40.00 with moderate non-cognitive attributes, yielding an overall performance score of 42.30, computed using Equation~\eqref{eq:performance}.

These performance scores were incorporated into the system prompt to adapt instructional responses dynamically. For Student~1, the AI generated a concise and structured explanation, assuming prior knowledge and emphasising key concepts with minimal scaffolding. This aligns with an advanced learner profile, where efficiency and higher-order reasoning are prioritised. Conversely, Student~2 received a more supportive and accessible explanation, employing simpler language, framing Python as beginner-friendly, and providing broader justification. This reflects increased scaffolding and reduced cognitive load appropriate for a lower performance level.

Although both students were presented with the same question, the AI-generated responses differed in tone, depth, and instructional strategy. This demonstrates that integrating cognitive and non-cognitive performance metrics enables adaptive and personalised feedback within the conversational learning pipeline.

The performance-based analysis of student interactions directly addresses \textbf{RQ4} by demonstrating that the proposed evaluation mechanism can effectively guide AI-generated explanations to better align with individual learner needs. These findings indicate that the system can dynamically adjust instructional support based on learner performance, reinforcing its effectiveness as an interactive, AI-driven learning and assessment environment.

\begin{table*}
\centering
\caption{Summary of Functional Evaluation Results and System Capabilities.}
\label{tab:functional_discussion}
\begin{tabularx}{\textwidth}{l X X}
\toprule
\textbf{Category} & \textbf{Key Results} & \textbf{Discussion and Implications} \\
\midrule
Authentication &
User registration and login functions operated correctly, with validation and error handling in place. &
This confirms the reliability of the authentication workflow and backend validation mechanisms, ensuring secure and user-friendly access control. \\
\addlinespace[0.5em]

Navigation and Content Delivery &
Core pages loaded correctly, enabling users to browse courses and access platform features. &
The results indicate stable routing and effective frontend–backend communication, supporting smooth navigation and content accessibility. \\
\addlinespace[0.5em]

Course and Content Management &
Institutions successfully created courses, modules, and submodules. \gls{rag} was employed to embed learning materials into the system, enabling the generation of up to 30 assessments and the integration of curated visual resources from YouTube. &
These results demonstrate robust backend orchestration of generative pipelines, supporting scalable and automated content authoring for educators. \\
\addlinespace[0.5em]

Learning and Assessment Interaction &
Students engaged with learning resources such as PDFs and generated videos, completed 10 randomly selected assessments, interacted with a chatbot grounded in course materials and LLMs, and received adaptive feedback based on performance. &
This confirms the system’s ability to support personalised and interactive learning experiences by integrating generative AI with pedagogical workflows. \\
\addlinespace[0.5em]

Dashboard and User Profile &
Students accessed dashboards that displayed learning progress and performance metrics, and updated their non-cognitive attributes through their profiles. &
The inclusion of performance analytics and non-cognitive data supports holistic learner modelling and enables adaptive learning and future personalisation strategies. \\

\bottomrule
\end{tabularx}
\end{table*}



\section{Conclusion}\label{sec8}
This paper set out to enhance educational access and learning outcomes in multilingual and under-resourced contexts through the design, development, and evaluation of an AI-enhanced personalised and adaptive learning platform. Guided by clearly defined research objectives, the study delivered a coherent end-to-end solution that integrates conversational AI, adaptive assessment, multilingual support, and learner analytics within a unified educational system.

A central contribution of this work is the demonstration of how \gls{llm}, fine-tuned for Nigerian Pidgin English, can be effectively applied to promote linguistic inclusivity and reduce barriers to learning in contexts where dominant instructional languages marginalise large segments of the learner population. By integrating \gls{rag}, the platform addresses the limitations of static curricula and hallucination-prone generative models, enabling the delivery of contextually grounded, up-to-date, and curriculum-aligned learning content. The inclusion of multimodal instructional resources further supports diverse learning preferences and enhances learner engagement.

From a technical and deployment perspective, the study provides empirical insights into the performance–efficiency trade-offs of quantised language models in low-resource environments. Through systematic evaluation using both automatic metrics and human-centred assessments, the results show that higher-capacity models achieve superior fluency, coherence, and semantic reliability, while lower-bit quantised models remain viable for deployment where computational resources are constrained. These findings offer practical guidance for selecting and configuring AI models in real-world educational settings with limited infrastructure.

The platform also advances adaptive learning practice by incorporating both cognitive performance indicators and selected non-cognitive learner attributes within a transparent, rule-based assessment framework. This hybrid approach enables interpretable and accountable adaptive decision-making, supporting personalised feedback, learning pathway adjustment, and learner self-awareness. Role-based analytics dashboards further provide educators and learners with actionable insights into progress, engagement, and learning challenges.

Despite these contributions, several limitations should be acknowledged. The evaluation was conducted within a controlled experimental setting, and large-scale longitudinal deployment in formal educational institutions was beyond the scope of this study. Learner engagement and learning outcomes were assessed primarily through short-term interactions and proxy measures, rather than extended academic performance over time. In addition, while Nigerian Pidgin English was selected as a representative low-resource language, further validation across additional indigenous languages would strengthen the generalisability of the approach.

Future work will focus on large-scale field deployment and longitudinal studies to evaluate learning gains, retention, and equity outcomes over extended periods. Planned enhancements include the integration of machine learning–based learner modelling to complement the rule-based framework, expansion to additional Nigerian and African languages, and the incorporation of offline and low-bandwidth functionality to improve accessibility in connectivity-limited environments. Further investigation into ethical considerations, data privacy, and responsible AI governance in educational contexts will also be prioritised.

In general, this research demonstrates the practical viability and educational value of AI-driven personalised and adaptive learning platforms tailored to multilingual and low-resource contexts. By combining advances in generative AI, adaptive assessment, and inclusive design, the proposed system offers a scalable and context-aware pathway toward improving accessibility, inclusivity, and learning outcomes in regions where educational inequities remain most pronounced.

\printcredits

\bibliography{cas-refs}
\end{document}